\documentclass{JHEP3}

\usepackage{amsmath}
\usepackage{latexsym}
\usepackage{graphicx}
\usepackage{mathrsfs}

\allowdisplaybreaks[1]

\numberwithin{equation}{section}

\def\AdSs5{$AdS_5$}
\def\AdSS5{$AdS_5$}
\def\AdS5s5{$AdS_5 \times S^5$}
\def\al{{\alpha^{\prime}}}
\def\gs{g_{st}}
\def\gy{g_{_{\rm YM}}}
\def\er{{\rm e}}
\def\dr{{\rm d}}
\def\Tr{{\rm Tr}}

\def\gs{g_{\rm s}}

\newcommand{\ie}{{\it i.e.~}}

\newcommand{\s}{\sigma}

\newcommand{\be}{\begin{equation}}
\newcommand{\ee}{\end{equation}}
\newcommand{\ba}{\begin{eqnarray}}
\newcommand{\ea}{\end{eqnarray}}
\newcommand{\bdm}{\begin{displaymath}}
\newcommand{\edm}{\end{displaymath}}
\newcommand{\ra}{\rangle}
\newcommand{\la}{\langle}
\newcommand{\pp}{\prime}

\newcommand\fr[1]{\frac{1}{#1}}

\newcommand{\wtil}{\widetilde}
\newcommand{\what}{\widehat}

\newcommand{\bea}{\begin{eqnarray}}
\newcommand{\eea}{\end{eqnarray}}%
\newbox\SlashedBox
\def\fs#1{\setbox\SlashedBox=\hbox{#1}
\hbox to
0pt{\hbox to 1\wd\SlashedBox{\hfil/\hfil}\hss}{#1}}
\def\hboxtosizeof#1#2{\setbox\SlashedBox=\hbox{#1}
\hbox to
1\wd\SlashedBox{#2}}

\def\ms#1{\setbox\SlashedBox=\hbox{$#1$}
\hbox to 0pt{\hbox to
1\wd\SlashedBox{\hfil/\hfil}\hss}#1}

\def\t2{\tau_2}
\def\IZ{\relax\ifmmode\mathchoice {\hbox{\cmss Z\kern-.4em Z}}
{\hbox{\cmss Z\kern-.4em Z}}
{\lower.9pt\hbox{\cmsss Z\kern-.4em Z}}
{\lower1.2pt\hbox{\cmsss Z\kern-.4em Z}}
\else{\cmss Z\kern-.4em Z}\fi}

\def\a{{\alpha}}
\def\g{\gamma}
\def\veps{\varepsilon}

\def\adot{{\dot\alpha}}

\def\D{\Delta}

\def\c1{{\chi^1}}

\def\v{\varphi}

\def\N4{{\cal N}=4}

\def\nn{\nonumber}

\def\nsix{(\bar\nu \nu)_{\bf 6}}

\newcommand{\calN}{{\mathcal N}}

\newcommand{\scrO}{{\mathscr O}}

\DeclareMathAlphabet{\mathpzc}{OT1}{pzc}{m}{it}

\newcommand{\scrm}{{\mathpzc m}}

\newcommand\hsp[1]{\hspace*{#1 cm}}

\newcommand{\mb}[1]{\mathbf{#1}}

\title{Non-Perturbative Contributions in the Plane-Wave/BMN Limit}
\author{Michael B. Green$^\dagger$, Stefano Kovacs$^{\#}$ and
  Aninda Sinha$^\dagger$ \\ {$^\dagger$ \it   Department of Applied
Mathematics and
Theoretical Physics,\\ Wilberforce Road, Cambridge CB3 0WA,
UK} \\ \email{M.B.Green, A.Sinha@damtp.cam.ac.uk} \\
$^{\#}$ Max-Planck-Institut f\"ur Gravitationsphysik \\
Albert-Einstein-Institut \\
Am M\"uhlenberg 1, D-14476 Golm, Germany  \\
\email{stefano.kovacs@aei.mpg.de}}

\abstract{This talk surveys recent work on the contribution of
instantons to the anomalous dimensions of  BMN operators in $\calN=4$
supersymmetric Yang--Mills theory and the corresponding
non-perturbative contributions to the mass-matrix of excited string
states in maximally supersymmetric plane-wave string theory.   The
dependence on the coupling constants and the impurity mode numbers in
the gauge theory and string theory are in striking agreement.

[Presented by MBG at the  Einstein Symposium, Bibliotecha  Alexandrina,
June 4--6 2005.]}

\preprint{DAMTP-2005-100 \\ AEI-2005-158}


\begin{document}

\section{Introduction}
\label{intro}

The conjectured correspondence between the BMN sector of ${\cal N}=4,
d=4$ supersymmetric
Yang-Mills  and type IIB string theory in the maximally
supersymmetric plane-wave background   has been examined in some
detail at the
perturbative level. However, the
understanding of non-perturbative
aspects of the correspondence has been very limited.
Such non-perturbative effects are well-studied
in the context of the AdS/CFT
correspondence where Yang--Mills instanton effects in $\calN =4$
supersymmetric
Yang-Mills correspond closely to $D$-instanton effects in type IIB
superstring theory  in $AdS_5\times S^5$. A
natural
question to ask is whether there is a similar relationship between
non-perturbative effects
in plane-wave string theory and the BMN limit of the gauge theory.

The correspondence relates the plane-wave string mass spectrum to the spectrum of
scaling dimensions of gauge theory operators in the so called BMN
sector of $\calN$=4 SYM. This consists of gauge invariant operators of
large conformal dimension, $\D$, and large charge, $J$, with respect
to a U(1) subgroup of the SU(4) R-symmetry group. The duality involves
the double limit $\D\to\infty$, $J\to\infty$, while $\D-J$ is kept
 finite and related to the string theory hamiltonian by
\be
\D-J = \fr{\mu} \, H^{(2)} \, ,
\label{dict}
\ee
The background value of the Ramond--Ramond ($R-R$) five-form flux, $\mu$, is
related to the mass parameter, $m$, which appears in the light cone
string action by $m=\mu p_-\al$, where $p_-$ is a component of
light cone momentum.
 The   two-particle hamiltonian is the sum of two pieces
\be
H^{(2)}=H^{(2)}_{\rm pert}+H^{(2)}_{\rm nonpert}\,.
\ee
The perturbative part, $H^{(2)}_{\rm pert}$, is a power series in the
string
coupling,  $g_s$, while  $H^{(2)}_{\rm nonpert}$ is the non-perturbative
part, which is suppressed by powers of $e^{-1/g_s}$.

The correspondence between the spectra of the two theories is the
statement that the eigenvalues of the operators on the two sides of
the equality (\ref{dict}) coincide.   A quantitative comparison is
possible if one considers the large $N$ limit in the gauge theory
focusing on operators in the BMN sector. As a result of combining the
large $N$ limit with the limit of large $\D$ and $J$, new effective
parameters arise, which are related to the ordinary
't Hooft parameters, $\lambda$ and $1/N$, by a rescaling,
\be
\lambda^\pp = \frac{\gy^2N}{J^2} \, , \qquad g_2 = \frac{J^2}{N} \, .
\label{g2lambpdef}
\ee
The correspondence relates these effective gauge theory couplings to
string theory parameters in the plane-wave background,
\be
m^2=(\mu p_-\al)^2 = \fr{\lambda^\pp} \, , \qquad
4\pi\gs m^2 = g_2 \, .
\label{paramid}
\ee
The double scaling limit, $N\to\infty$, $J\to\infty$, with $J^2/N$
fixed, connects the weak coupling regime of the gauge theory to string
theory at small $g_s$ and large $m$.

Perturbative contributions to the mass spectrum have been
analysed in some detail on both the string side and compared with corresponding
contributions to the anomalous dimensions of BMN operators in the gauge theory.
However,  there have been no calculations of
non-perturbative corrections due to $D$-instanton effects, which
contribute to  $H^{(2)}_{\rm nonpert}$ or of the corresponding Yang--Mills instanton
contributions in the BMN limit.  Indeed, it is not at all obvious at first
sight that Yang--Mills instantons survive the BMN limit, but the correspondence with
string theory $D$-instantons implies that they must.   This talk, which is necessarily brief,
reviews the contents of
\cite{gks1,gks2} that study the BMN/plane-wave correspondence at the
non-perturbative level and details can be found in these papers\footnote{No further
bibliographic references will be given here but full bibliographies are
contained in these publications.}.
The next section summarizes the results  of \cite{gks1} on
plane-wave string theory while the gauge theory results of
\cite{gks2} are summarized in section 3.  The agreement between the dependence of
the instanton contributions on the two sides of the correspondence is impressive.
Further details concerning states with fermionic impurities are in a forthcoming publication
\cite{gks3}.

\section{Mass-matrix elements in plane-wave string theory}
In the maximally supersymmetric
plane-wave background the five-form $R-R$ potential has a non-zero value
that sets the scale for the masses of the supergravity fields and reduces the
isometry of the background to
SO(4)$\times$ SO(4).
Light-cone gauge string theory in this background is a free
world-sheet theory with eight massive world-sheet bosons, $X^I$, and eight massive world-sheet
fermions, $\theta^A$, and may be  described by the world-sheet action
\be
\label{wsaction}
{\cal L}= {1\over 2}\left(\partial_+X\, \partial_- X - m^2
  X_I^2\right) + i \left(\theta^1 \partial_+  \theta^1 +
\theta^2 \partial_-  \theta^2 - 2m\,\theta^1\,\Pi\,\theta^2\right)\, ,
\ee
where $\theta^1$ and  $\theta^2$ are SO(8) Grassmann spinors and $\Pi = \gamma^1\gamma^2\gamma^3\gamma^4$
(where $\gamma^I$ are SO(8) gamma matrices).
In the quantum theory the zero modes in the eight transverse
directions, $X^I$, define harmonic oscillators with strength proportional to $m$.

The classical supergravity states are obtained by applying the zero mode bosonic and fermionic
creation operators to the ground state (the BMN vacuum).  Excited string states are constructed
as usual by applying higher mode creation operators to the zero mode states.  A state constructed by
applying $p$ excited bosonic or fermionic creation operators is said to have $p$ `impurities', a
terminology that makes contact with the corresponding operators in the gauge theory.  Each oscillator
can be in any excited state subject to the usual `level-matching'  restriction which means that there
are, in general, $p-1$ independent mode numbers that enter into the definition of the $p$-impurity
state.
Some effort has been expended in constructing a three-string vertex, from which a certain
amount of perturbative information concerning string two-point functions -- or,
equivalently, the mass matrix elements --  beyond free string theory can be extracted.
We are concerned with non-perturbative contributions to the hamiltonian due to $D$-instantons.
The single $D$-instanton sector  has a measure that is proportional to $e^{2i\pi \tau}$
where $\tau= \tau_1 + i \tau_2 \equiv C^{(0)} + i e^{-\phi}$
($C^{(0)}$ is the $R-R$ pseudoscalar, $\phi$ is the dilaton
and $g_s= e^{\phi}$).
Although this is exponentially small, it is the
leading contribution  with the phase factor $e^{2\pi i C^{(0)}}$.  It
is therefore of  interest to understand how the mass matrix is
modified by these contributions.  In the following we will outline the
calculation of  such $D$-instanton contributions to mass matrix elements, or two-point functions,
to leading order in the string coupling.

A $D$-instanton with position   $x_0$
is described, to lowest order in the string coupling, by world-sheet
disks with Dirichlet boundaries fixed at $x_0$.   The light-cone
 boundary state description of the $D$-instanton in plane-wave string theory,
  generalizes that  of the Minkowski space theory.
The $D$-instanton boundary state couples to single closed-string states and preserves eight
kinematical and eight dynamical supersymmetries and is given (at a specific value of $x_0^+$)
 by
\be
|B, {\bf x_0}\rangle = {\cal N}_{(0,0)}
\exp \left( \sum_{k=1}^{\infty}
\frac{1}{\omega_k} \alpha^I_{-k} \tilde\alpha^I_{-k}
- i {\eta}  S_{-k}M_k\tilde{S}_{-k}\right) \,
|  {\bf x_0}\,\rangle_0\,,
\label{boundone}
\ee
where $\alpha$, $\tilde \alpha$, $S$ and $\tilde S$ are the left and right-moving
non-zero modes of the bosonic and fermionic coordinates, $X^I$, $\theta^1$ and $\theta^2$, and
$| {\bf x_0}\,\rangle_0$ is the ground state of all
the oscillators of non-zero mode number. The coordinate ${\bf x_0}^I$
is the eigenvalue of the position operator
constructed from the zero-mode oscillators, $a^{\dagger\, I}$ and $ a^I$.
The quantity $M_k$ is a matrix in spinor space and is a function of $m$ that reduces to
the unit matrix in the flat-space limit, and $\omega_{k}=\sqrt{m^2+ k^2}$.

The leading contribution to the two-point function of string states in
a $D$-instanton background comes from a disconnected
world-sheet that is the product of two disks, with one closed-string
state attached to each, and with Dirichlet boundary conditions.  The two-boundary state is
simply the product of two single-boundary states acting in distinct Fock spaces,
$||B\, {\bf x_0}\rangle\rangle_2 =|B, {\bf x_0}\rangle \otimes |B, {\bf x_0}\rangle'$.
The position coordinates $x_0$ are interpreted as bosonic
moduli associated with broken translation invariance.
Similarly, the eight broken kinematical supersymmetries, $q^a$ ($a=1,\dots,8$), and
the eight broken dynamical supersymmetries, $Q^{\dot a}$ ($\dot a=1,\dots,8$),
lead to the presence of sixteen fermionic
moduli, $\epsilon^a$, $\eta^{\dot a}$.
 The dependence of the state on the fermionic moduli is included by summing over all
possible ways of attaching sixteen fermionic open-string states to the boundaries of
the two disks. A dressed boundary state, including these supermoduli can be defined by
\be
\label{dressvert}
||V, x_0\rangle\rangle_2 = g_s^{7/2}\,e^{2\pi i \tau}\,
\prod_{a=1}^8(\epsilon^a q^a)\, (\eta^aQ^a)\, ||B\, {\bf x_0}\rangle\rangle_2
 \, e^{i  x_0^+ (p_{1+} + p_{2+})} \, e^{i x_0^- (p_{1-} + p_{2-})}\,.
\ee
 The factor
of $g_s^{7/2}$ in the $D$-instanton measure can be extracted from previous work on
$D$-instanton
contributions in $AdS_5\times S^5$ (and we are not keeping overall
multiplicative constants).
The on-shell two-point function between string states
$|\chi_1\rangle$ and $\langle \chi_2|$
is given by the integrated matrix element
\be
\label{matel}
\int d^8 \epsilon\, d^8 \eta\, d^{10}x_0\,
\langle \chi_2|\otimes\langle \chi_1||V,x_0\rangle\rangle_2\, .
\ee
Integration over the light-cone moduli, $x_0^\pm$, ensures the conservation of $p_\pm$ in
any process while integration over the other supermoduli generate correlations between the
two disks.

For a state with occupation numbers ${n_{r}}$ (where $r$ labels the
oscillator levels) the light-cone energy
 is given by  the nonlinear formula $p_+ = \sum_{r} \omega_r/2p_-$.
It follows that conservation of $p_+$ implies that
the number of impurities is preserved by this process, so that
$|\chi_1\rangle$ and $\langle \chi_2|$
have the same number of impurities.  Generally,  conservation
of $p_+$ imposes the even stronger condition that the non-zero
mode numbers of oscillators in the incoming state coincide with those
of the outgoing state. The nonlinear energy relation
is seen on the gauge side after summing
perturbative planar contributions to all orders in $\lambda'$.  However,
to leading order in the
$1/m^2\sim \lambda'$ expansion $\omega_{n_r} = m$ and conservation of
$p_+$ imposes no relation between the mode numbers of the
incoming and outgoing states. Therefore, since we are interested in comparing with
perturbative gauge theory we need not impose the equality of incoming and
outgoing oscillator mode numbers in the following.

Certain other general features of
$D$-instanton dominated matrix
elements follow from general properties of the boundary state (\ref{dressvert}).
For example, the boundary state couples to
arbitrary numbers of pairs of modes, where each pair consists of one
left-moving mode with a mode-number $n$ and a right-moving mode with the {\it same} mode number.
This means that it only has non-zero coupling to states that are level-matched in this pairwise fashion
-- a feature that must therefore also be true on the gauge theory side although it will prove
 much harder to see  this from a conventional Yang--Mills instanton calculation.

Examples of $D$-instanton contributions to  matrix elements between states with various numbers of
bosonic and fermionic impurities were considered in \cite{gks1}.
Those results that are particularly relevant for comparison
with the gauge theory results of \cite{gks2} are the following.

\medskip
{\it  Two bosonic impurities}

A level-matched state with  two bosonic impurities is associated with  a single mode number.
The two-state bra vector in (\ref{matel}) is given by
\be
 {_{{}_1}}\langle \chi_1|\otimes {_{{}_2}}\langle\chi_2| =
 \frac{1}{\omega_m \omega_n}\, t_{IJ}^{(1)}\, t_{KL}^{(2)}
 { }_h\langle 0|\alpha_m^{(1) I}\, \tilde \alpha_m^{(1) J}
 \otimes  {}_h\langle 0|\alpha_n^{(2)K}\, \tilde \alpha_n^{(2)L}  \, ,
\label{bostates}
\ee
where the wave functions $t_{IJ}^{(1)}$ and
$t^{(2)}_{IJ}$ are tensors of SO(4)$\times$ SO(4)
(with indices that take the values $I,J,K,L=1,\dots,8$).
The ground state $|0\rangle_h$ denotes the BMN ground state, which is the
state of lowest $p_+$.
The leading semi-classical one $D$-instanton contribution to the two-string mass-matrix element
is  independent of the mode number and has the form (ignoring a constant overall factor)
\bea
\label{twoimp}
{1\over \mu}H^{(2)}_{\rm nonpert} &=&
e^{2\pi i \tau}\, g_s^{7/2}\, m^3\, t^{(1)}_{ij}\, t^{(2)}_{pq}\,
(\delta^{pi}\delta^{qj}+\delta^{pj}\delta^{iq})\nn\\
&=&e^{i\theta-{8\pi^2\over g_{_{YM}}}}\, {\lambda^{\prime}}^2\, g_2^{7/2}
\, t^{(1)}_{ij}\, t^{(2)}_{pq}\,
(\delta^{pi}\delta^{qj}+\delta^{pj}\delta^{iq})
\, ,
\eea
 in the large-$m$ limit (where $m=\mu\, p_-\,\alpha'$). We have indicated the expression
 in terms of the gauge theory parameters in the second line for future reference.
 In this expression we have also  specialized to
 vector indices $i,j,p,q$ lying  in one of the SO(4) factors of the
 SO(4)$\times$ SO(4) isometry  group since this is the case that is easiest to calculate
 in the gauge theory.
Although the exact string theory expression includes all non-leading terms, it is
only the large-$m$ limit that can be compared with the gauge theory calculations.
Note, in particular,  that this leading contribution is of order ${\lambda'}^2$ and is suppressed
relative to potential $O({\lambda'}^0)$ effects.  This
fact makes the Yang--Mills instanton contribution to the two impurity
case more difficult to evaluate in precise detail than cases with
higher numbers of impurities.

\medskip

{\it Four bosonic impurities}

With four bosonic impurities there are three independent
  non-zero mode numbers for each external state after taking level
  matching into account.  However, as we remarked above, the only non-zero matrix elements
  are those in which each
$\alpha_n$ mode is accompanied by a  $\tilde \alpha_n$ with the same
mode number, $n$.
 In this case the bra state in (\ref{matel}) is given
by
\bea
&& {_{{}_1}}\langle \chi_1|\otimes {_{{}_2}}\langle\chi_2| =
\frac{1}{\omega_{m_1} \omega_{m_2}\omega_{n_1} \omega_{n_2}}\, t_{j_1j_2j_3j_4}^{(1)}\,
t_{p_1p_2p_3p_4}^{(2)}\nn\\
 && { }_h\langle 0|\alpha_{m_1}^{(1) j_1}\, \tilde \alpha_{m_1}^{(1) j_2}\,
 \alpha_{m_2}^{(1) j_3}\, \tilde \alpha_{m_2}^{(1) j_4}
 \otimes  {}_h\langle 0|\alpha_{n_1}^{(2)p_1}\, \tilde \alpha_{n_1}^{(2)p_2}  \,
  \alpha_{n_2}^{(2) p_3}\, \tilde \alpha_{n_2}^{(2) p_4}\, .
\label{bostates2}
\ea
The tensor wave functions  $t_{p_1p_2p_3p_4}$
of the in and out states  have again been restricted to
have indices in a single SO(4) factor of the isometry group simply
because that is the easiest case to consider in the dual gauge
theory.  In this case the mass-matrix is given, at leading order in
powers of $1/m$, by
\ba
{1\over \mu}H^{(2)}_{\rm nonpert}&=&
 e^{2\pi i \tau}\,t^+\, t^-\, g_s^{7/2}\, m^7\,
{1\over m_1m_2n_1n_2}\nn\\
&=&  e^{i\theta-{8\pi^2\over g_{_{YM}}}}\, t^+\, t^-\,g_2^{7/2}\,
{1\over m_1m_2n_1n_2}\, ,
\label{finfour}
\ea
where
\be
t^\pm =  t_{j_1\tilde j_2j_3\tilde j_4}\,
(\delta_{j_1j_3}\delta_{j_2
  j_4}-\delta_{j_1 j_4}\delta_{j_2 j_3}\pm\epsilon_{j_1 j_2 j_3 j_4})\, .
\label{tpmdef}
\ee
In this case the result is zeroth order in $\lambda'$ perturbation
theory.  The expression (\ref{finfour}) implies that to leading order
in $m$ only scalar states  have an induced $D$-instanton coupling.
The rest of the possible
bosonic four-impurity states have couplings that are
 suppressed by powers of $m$ compared to this
leading result.  Further details of these four-impurity matrix elements are given in
\cite{gks1}.

\section{Anomalous dimensions of BMN states in $\calN=4$   Yang--Mills theory}

We will now discuss semi-classical instanton contributions
to the anomalous dimensions of BMN operators  in $\calN=4$ SU(N) Yang--Mills
theory, which are extracted from two-point correlation functions.

Conformal invariance determines the form
of two-point functions of primary operators, $\scrO$ and $\bar\scrO$,
to be
\be
\la\scrO(x_1)\bar\scrO(x_2)\ra = \frac{c}{(x_1-x_2)^{2\D}} \, ,
\label{conf2pt}
\ee
where $\D$ is the scaling dimension. In general in the quantum theory
$\D$ acquires an anomalous term, $\D(\gy)=\D_0+\g(\gy)$. At weak
coupling the anomalous dimension $\g(\gy)$ is small and substituting in
(\ref{conf2pt}) gives
\be
\la\scrO(x_1)\bar\scrO(x_2)\ra
=\frac{c\,\Lambda^{2\g(\gy)}}{(x_1-x_2)^{2\D_0}}
\left( 1-\g(\gy) \log\left[\Lambda^2(x_1-x_2)^2\right] + \cdots
\right) \, ,
\label{adimexp}
\ee
where $\Lambda$ is an arbitrary renormalisation scale. As a function of
the coupling constant the anomalous dimension admits an expansion
consisting of a perturbative series plus non-perturbative
corrections. The generic two-point function at weak coupling takes the
form
\ba
\la\scrO(x_1)\bar\scrO(x_2)\ra &=& \frac{c(\gy)}{(x_1-x_2)^{2\D_0}}
\left( 1-\gy^2\g^{(1)} \log\left[\Lambda^2(x_1-x_2)^2 \right] \right.
\nn \\
&& \left. + \cdots - \er^{2\pi i\tau}\g^{\rm (inst)}
\log\left[\Lambda^2(x_1-x_2)^2\right] + \cdots \right) \, .
\label{adimexp2}
\ea
Therefore perturbative and instanton contributions to the anomalous
dimension are extracted from the coefficients of the logarithmically
divergent terms in a two-point function.
The general  structure of  these anomalous dimensions  is an expansion of the form
\be
\g(\gy,\theta, N) = \sum_{n=1}^\infty \g^{\rm pert}_n(N)\,\gy^{2n} +
\sum_{K>0}\sum_{m=0}^\infty \left[\g^{(K)}_m(N)\,\gy^{2m}\,
\er^{2\pi i\tau K} +\;\mathrm{c.c.} \, \right] \, ,
\label{gamma-exp}
\ee
where $\tau = \frac{\theta}{2\pi}+i\frac{4\pi}{\gy^2}$. The double
series in the second term in (\ref{gamma-exp}) contains the
contributions of multi-instanton sectors as well as the perturbative
fluctuations in each such sector.
If the BMN sector of the gauge theory scales appropriately
 (\ref{gamma-exp}) becomes a
series in the scaled couplings $\lambda^\pp$ and $g_2$, as we will
see in the examples to be reviewed in this section.

The gauge invariant composite BMN operators are defined by
normalized traces of products of scalar fields and are
labelled by the total value of $\D -J$, the impurity number.
The six scalar fields in the Yang--Mills multiplet are decomposed
according to the value of $J$ they carry:  $Z$ has $J=1$ or $\D-J=0$, $\bar Z$ has $J=-1$
or $\D+J=0$ and $\v^i$ ($i=
1,2,3,4)$ have $J=0$ or $\D\pm J=1$.   Since we want to take the limit of large $J$
for a fixed value of $\D-J$ a BMN operator has a large number of $Z$'s and a finite number
of other scalar impurities.
 Operators with $\D-J=0$ or $1$ are protected and so their
two-point functions do not receive instanton contributions.
With two or more impurities the situation is more interesting although the analysis is
quite complicated. We are considering BMN operators with $k$ bosonic impurities
that are linear combinations of colour traces of products of scalar fields of the form
\ba
\scrO^{i_1\ldots i_k}_{J;n_1\ldots n_k}
&\!=\!& \fr{\sqrt{J^{k-1}\left(\frac{\gy^2N}{8\pi^2}\right)^{J+k}}}
\hspace*{-0.1cm} \begin{array}[t]{c}
{\displaystyle \sum_{p_1,\ldots,p_{k-1}=0}^J} \\
{\scriptstyle p_1+\cdots+p_{k-1} \le J}
\end{array} \hspace*{-0.1cm} \er^{2\pi i[(n_1+\cdots+n_{k-1})p_1+
(n_2+\cdots+n_{k-1})q_3+\cdots+n_{k-1}p_{k-1}]/J} \nn \\
&& \hsp{3} \times \Tr\left(Z^{J-(p_1+\cdots+p_{k-1})}\v^{i_1}Z^{p_1}
\v^{i_2}\cdots Z^{p_{k-1}}\v^{i_k}\right) \, .
\label{k-scal-imp}
\ea
The conjugate operator $\bar\scrO$ has a large number of
$\bar Z$'s instead of $Z$'s.  For the two-point function (\ref{adimexp2})  to
be non-vanishing
the operators must have equal and opposite values of $J$.

\medskip
{\it Instanton contributions to two-point correlation functions}

 In semi-classical approximation, correlation functions of composite
operators are computed by replacing each field by the solution of its
corresponding field equation in the presence of an instanton,
expressed in terms of the fermionic and bosonic moduli.  These moduli
encode the broken superconformal symmetries together with the broken
(super)symmetries associated with the orientation of a SU($2$)
instanton within SU($N$).  The symmetries are restored by integration over
the supermoduli.  For large $N$, the integration is carried out by a saddle point
procedure.  In the case of a
two-point function of a generic local operator, $\scrO(x)$, and its
conjugate the supermoduli integral  takes the form
\be
\la\bar\scrO(x_1)\scrO(x_2)\ra_{\rm inst} = \int \dr \mu_{\rm inst}
(\scrm_{\rm \,b},\scrm_{\rm \,f}) \, \er^{-S_{\rm inst}} \;\,
\hat{\bar{\!\!\scrO}}(x_1;\scrm_{\rm \,b},\scrm_{\rm \,f})
\hat\scrO(x_2;\scrm_{\rm \,b},\scrm_{\rm \,f})\, ,
\label{semiclass}
\ee
where we have denoted the bosonic and fermionic collective coordinates
by $\scrm_{\rm \,b}$ and $\scrm_{\rm \,f}$ respectively. In
(\ref{semiclass}) $\dr\mu_{\rm inst}(\scrm_{\rm \,b}, \scrm_{\rm\,f})$
is the integration measure on the instanton moduli space, $S_{\rm
inst}$ is the classical action evaluated on the instanton solution and
$\hat\scrO$ and $\hat{\bar{\!\!\scrO}}$ denote the classical
expressions for the operators $\scrO$ and $\bar\scrO$ computed in the
instanton background.

A one-instanton configuration in SU($N$) Yang--Mills theory is
characterised by  $4N$ bosonic moduli that can be
identified with the size, $\rho$, and position, $x_0$, of the
instanton as well as its global gauge orientation. The latter can be
described by three angles identifying the iso-orientation of a SU(2)
instanton and 4$N$ additional constrained variables, $w_{u\adot}$ and
$\bar w^{\adot u}$ (where $u=1,\ldots,N$ is a colour index), in the
coset SU($N$)/(SU($N-2$)$\times$U(1)) describing the embedding of the
SU(2) configuration into SU($N$). In the one-instanton sector in the
$\calN$=4 theory there are additionally 8$N$ fermionic collective
coordinates corresponding to zero modes of the Dirac operator in the
background of an instanton. They comprise the 16 moduli associated
with Poincar\'e and special supersymmetries broken by the instanton
and denoted respectively by $\eta^A_\a$ and $\bar\xi^{\adot A}$ (where
$A$ is an index in the fundamental of the SU(4) R-symmetry group) and
8$N$ additional parameters, $\nu^A_u$ and $\bar\nu^{Au}$, which can be
considered as the fermionic superpartners of the gauge orientation
parameters.  The sixteen superconformal moduli are exact, \ie they
enter the expectation values (\ref{semiclass}) only through the
classical profiles of the operators. The other fermion modes,
$\nu^A_u$ and $\bar\nu^{Au}$, appear explicitly in the integration
measure via the classical action, $S_{\rm inst}$, and are therefore `non-exact'
moduli. This distinction
plays a crucial r\^ole in the calculation of correlation functions.
The $\nu^A_u$ and $\bar\nu^{Au}$ modes satisfy the fermionic ADHM
constraints
\be \bar w^{\adot u}\nu^A_u = 0 \, , \quad
\bar\nu^{Au}w_{u\adot} = 0 \, ,
\label{nuconstraint}
\ee
which effectively reduce their number to $8(N-2)$.
The manner in which these moduli enter into the expressions for the
fields is determined by the solution of the  field equations for
$\calN$=4 SYM theory in an instanton
background.  The solution for each field in the Yang--Mills suprermultiplet
 can be written as a sum of terms containing different numbers of
fermionic zero modes.  For the purpose of this talk  let us note that
a scalar field has the form
\ba
&&
\Phi^{AB} = \hspace*{-0.3cm} \begin{array}[t]{c}
{\displaystyle \sum_{n=0}} \\
{\scriptstyle 4n+2 \le 8N}
\end{array} \hspace*{-0.3cm}
\Phi^{(2+4n)AB} \, ,
\label{N4multizeromod}
\ea
where the notation $\Phi^{(4n+2)AB}$   denotes a term in the
solution for the field $\Phi$ containing a product of $4n+2$ fermion zero modes.  The minimum
number of fermionic moduli in a scalar field is therefore two, while the next
term contains a product of six fermionic moduli and so on.  It
is  understood that the number of
superconformal modes in each field cannot exceed 16 and the
remaining modes are of $\nu^A_u$ and $\bar\nu^{Au}$ type.  Furthermore, terms with higher
numbers of moduli are suppressed by powers of the coupling, so the leading contribution
to the two-point function is that with the minimal number of moduli in each scalar field.

In order to evaluate the two-point function
(\ref{semiclass}) the expressions for the fields in terms of moduli
must be substituted into each composite operator and the resulting
 traces must then be evaluated. The actual integration over the large number of
supermoduli is reasonably straightforward, but
there are complicated combinatorics involved in distributing the moduli
among the fields in the two operators, that we will now outline
(and are discussed in detail in \cite{gks2}).

The $J+k$ scalar fields in the operator $\scrO$ defined in (\ref{semiclass})
each contain at least two fermionic moduli, which may be chosen from the
superconformal moduli, $\eta$ and $\bar \xi$,
 or from the non-exact moduli, $\nu$ and $\bar \nu$.
The sixteen fermionic superconformal moduli naturally arise in the combination
\be
\zeta^A_\a(x) = \frac{1}{\sqrt{\rho}}\left[ \rho\,\eta^A_\a -
(x-x_0)_\mu \s^\mu_{\a\adot} \,\bar\xi^{\adot A} \right] \, ,
\label{zetadef}
\ee
where $\zeta^A_\alpha(x)$ are eight position-dependent Grassmann variables.
This means that there has to be a factor of $\prod_{A=1}^4(\zeta^A(x_1))^2$
in each operator in the two-point correlation function.  In other words each
of the two operators in the correlation function
has to contain eight of the superconformal moduli.
Taking their SU(4) quantum numbers into account
 only four of these can be soaked up by the $Z$ fields and the rest have to
 be contained in the impurity fields, $\v^{AB}$.
Once the sixteen superconformal moduli are distributed among some of the scalar fields
 the non-exact moduli are soaked up by the remaining (large number) of fields, which are
 mostly $Z$'s.

The bosonic integrations over the position and size of the instanton
are left as a last step. These
integrals are logarithmically divergent, the coefficient of the logarithm
corresponding to the contribution to the matrix of anomalous dimensions.

In \cite{gks2} we considered the two impurity and four-impurity cases in detail.
The results were as follows.

\medskip

{\it Two bosonic impurities}

For the two impurity case there is a technical problem in carrying out
a complete analysis.  The point is that in order to soak up all
sixteen of the fermionic supermoduli, at least one of the
scalars in each operator has to soak up six fermionic moduli, rather
than the minimum number of two.  This means that the contribution is
of higher order in $\lambda'$ than a leading contribution would be,
which is in line with the two-impurity result in plane-wave string
theory described earlier.  It is technically very complicated to derive
the precise from of this six-fermion contribution, but this  is needed to
determine the $J$-dependence of the two-point function.
Nevertheless, if we {\it assume} BMN scaling
the analysis can be carried through sufficiently
to argue that the result is in agreement with the string
calculation.  This follows since the
dependence on $\gy$ and $N$ can be determined without knowledge of the
details of the six-fermion term, and this
uniquely fixes the power of $J$ needed for BMN
scaling.  This requirement, in turn, constrains the way in which the
fermion zero modes can appear in the profile of the operator.
Specifically, the two-point function can obey BMN scaling only if the
distribution of the zero modes is such  that the final result  is
independent of the single mode number entering the definition of the
two impurity operators.

Since in this case the analysis is incomplete
we will only state the final result here, but will give a more detailed
description of our method in the four-impurity case.
It is simplest to choose the two states to be in the representation
$\mb9$ of SO(4)$_R$, since this sector contains only one operator
which cannot mix with any other.  The result yfor the two-point
function of this operator, assuming BMN scaling, has the form
\be
G_\mb9(x_1,x_2) \sim \frac{\gy^4J^3\er^{2\pi i\tau}}{N^{3/2}}
\fr{(x_1-x_2)^{2(J+2)}} \, I \, ,
\label{2pt-9-fin}
\ee
where $I$ is a logarithmically divergent integral over the bosonic
moduli, which can be  regulated by dimensional regularisation of the
$x_0$ integral. The coefficient of this divergence gives the instanton
induced anomalous dimension of $\scrO_\mb9^{\{13\}}$,
\be
\g_\mb9^{\rm inst} \sim \frac{\gy^4 J^3}{N^{3/2}}
\,\er^{-\frac{8\pi^2}{\gy^2}+i\theta} \sim \left(g_2\right)^{7/2}
\left(\lambda^\pp\right)^2 \
\er^{-\frac{8\pi^2}{g_2\lambda^\pp}+i\theta} \, .
\label{9-adim}
\ee
This is in agreement with the non-perturbative correction to the mass
of the dual string state computed in \cite{gks1}. In particular, the
anomalous dimension (\ref{9-adim}) is independent of the parameter $n$
corresponding to the mode number of the plane-wave string state.
Apart from the exponential factor characteristic of instanton effects,
(\ref{9-adim}) contains an additional factor of
$\left(\lambda^\pp\right)^2$. This is due to the inclusion of
six-fermion scalars which give rise to additional $\nsix$ bilinears,
each of which brings one more power of $\gy$. As will be seen in the
next subsection, in the case of four impurity SO(4)$_R$ singlets it is
sufficient to consider the solution for all the scalars that is
bilinear in the fermions and as a consequence we shall find a leading
contribution of order $(g_2)^{7/2}\er^{-8\pi^2/g_2\lambda^\pp}$.

\medskip

{\it Four bosonic impurities}

The calculation of two-point functions of four impurity operators is
more involved than the corresponding calculation in the two impurity
case from the point of view of the combinatorial analysis.
However, at the four impurity level, in the case of
SO(4)$_R\times$SO(4)$_C$ singlets, the calculation of the leading
instanton contributions requires only the inclusion
of the quadratic fermionic terms in the classical
profiles of the scalar fields, which
are known explicitly.

However, at
the four impurity level, in the case of SO(4)$_R\times$SO(4)$_C$
singlets, the
leading instanton contributions the classical profiles of the scalar fields
 involve only the quadratic fermionic terms and are known explicitly.
Therefore, in this case the semi-classical contributions to the two-point
functions can be analyzed more completely.
The fact that
non-zero correlation functions  are obtained using
the minimal number of fermion modes for each field also implies that
in this case a contribution to the matrix of anomalous dimensions arises at
leading order in $\lambda'$.   The case in which the external state
is an SO(4)$\times$ SO(4) singlet with four scalar impurities is the simplest to analyze
and also corresponds to the states we discussed in the context of the
plane-wave string theory.  More explicitly, the operators to be considered are
of the form
\ba
\scrO_{\mb1;J;n_1,n_2,n_3} &=&
\frac{\veps_{ijkl}}{\sqrt{J^3\left(\frac{\gy^2N}{8\pi^2}\right)^{J+4}}}
\begin{array}[t]{c}
{\displaystyle \sum_{q,r,s=0}^J} \\
{\scriptstyle q+r+s \le J}
\end{array} \hsp{-0.1} \er^{2\pi i[(n_1+n_2+n_3)q
+(n_2+n_3)r+n_3s]/J} \nn \\
&& \hsp{3} \times \Tr\left(Z^{J-(q+r+s)}\v^i
Z^{q}\v^jZ^{r}\v^kZ^{s}\v^l\right) \, ,
\label{epsdef}
\ea
which is dual to the scalar plane-wave string state
$\veps_{ijkl} \, \a^i_{-n_1}\a^j_{-n_2}\wtil\a^k_{-n_3}
\wtil\a^l_{-(n_1+n_2-n_3)} |0\ra_h$.   The conjugate operator involves
$\bar Z$'s instead of $Z$'s.

As before, in considering the distribution of the fermionic moduli
among the $J+4$ fields within a trace, half of the superconformal modes (i.e., eight)
must be
soaked up by each of the two operators in the two-point correlation function.
Furthermore, at least four of these have to be soaked up  by the impurity scalar fields
since the quantum numbers of the $Z$'s are such that they can soak up at most four of the
superconformal modes.  The number of possible ways of distributing  each kind of fermionic
modulus among the $J+4$ scalar fields is very large and we will not describe the combinatorics
here.   After summing this very large number of terms the resulting expression for the
correlator is  (omitting overall coefficients)
\ba
G_\mb1(x_1,x_2) &=& \frac{\er^{2\pi i\tau}}{J^32^JN^{7/2}}
\int \frac{\dr^4x_0\,\dr\rho}{\rho^5} \,
\frac{\rho^{J+8}}{[(x_1-x_0)^2+\rho^2]^{J+8}}
\frac{\rho^{J+8}}{[(x_2-x_0)^2+\rho^2]^{J+8}} \nn \\
&& \times \int \prod_{A=1}^4\dr^2\eta^A\dr^2\bar\xi^A \,
\prod_{B=1}^4 \left[\left(\zeta^B\right)^2(x_1)\right]
\left[\left(\zeta^B\right)^2(x_2)\right] \nn \\
&& \times \int \dr^5\Omega \, \left(\Omega^{14}\right)^J
\left(\Omega^{23}\right)^J \, K(n_1,n_2,n_3;J) K(m_1,m_2,m_3;J) \, ,
\label{2pt-4imp-1}
\ea
where $\Omega^{AB}$ are angular variables on the five-sphere that emerge
from the integration over the $\nu$ and $\bar\nu$ moduli.
The $J$ and $N$ dependence in the prefactor
is obtained combining the normalisation of the
operators, the contribution of the measure on the instanton moduli
space and the factors of $\gy\sqrt{N}$ associated with the $\nu$ and $\bar\nu$
variables.   The expression (\ref{2pt-4imp-1}) contains integrations over
the bosonic moduli, $x_0$ and $\rho$, the sixteen superconformal
fermion modes and the five-sphere coordinates $\Omega^{AB}$.

The
dependence on the integers $n_i$, $m_i$, $i=1,2,3$, dual to the mode
numbers of the corresponding string states is contained in the functions
$K(n_1,n_2,n_3;J)$ and $K(m_1,m_2,m_3;J)$. These are given by the sum
of 35 terms, which are sums over integers
$q,r,s$ of phases $\exp
\{2\pi i[(n_1+n_2+n_3)q+(n_2+n_3)r+n_3s]/J\}$ multiplying the multiplicity
factors associated with the different distributions of $\what Z$'s in
each case.  There are very many contributions to each of these 35 terms
and the sums over this very large number of phase factors lead to some
very impressive cancellations of what would otherwise be large and unlikely
looking expressions.

The final result is obtained after performing the bosonic integrals.
 At each step various powers of $g_{_{YM}}$, $N$ and $J$ enter,
 and it appears rather miraculous that in the end they all combine into a
 function that depends only on $g_2$ and $\lambda'$, in accord with the
 BMN scaling.  We can indicate where these different powers of the couplings
 come from as follows,
\ba
&& \underbrace{\left(\fr{\sqrt{J^3
(\gy^2N)^{J+4}}}\right)^2}_{\rm normalised ~ op.~ profile}
\;\underbrace{\left(\gy\sqrt{N}\right)^{2J}}_{\nu,\:\bar\nu
~ {\rm bilinears}}\;
\underbrace{\frac{\er^{2\pi i\tau}\gy^8\sqrt{N}}{2^J}}_{\rm measure}
\;\underbrace{\frac{2^J}{J^2}}_{S^5 ~ {\rm integral}}
\;\underbrace{\fr{J^2}}_{x_0,\:\rho ~ {\rm integrals}}
\;\underbrace{(J^7)^2}_{{\rm sums}} \nn \\
&& \sim \frac{J^7}{N^{7/2}}\, \er^{2\pi i\tau}=
(g_2)^{7/2}\,\er^{-\frac{8\pi^2}{g_2\lambda^\pp}+i\theta} \, .
\label{gNJ-dep}
\ea

The final result for the two-point function turns out to vanish unless the mode numbers of the
operators are equal in pairs -- just as in the string theory $D$-instanton calculation.
The result is
\be
G_1(x_1,x_2) = \frac{3^2(g_2)^{7/2}
\er^{-\frac{8\pi^2}{g_2\lambda^\pp}+i\theta}}{2^{41}\pi^{13/2}}\,
\fr{m_1m_2n_1n_2 } \fr{(x_{12}^2)^{J+4}} \,
\log\left(\Lambda^2x_{12}^2\right)\, ,
\label{4imp-2pt-as}
\ee
where the scale $\Lambda$ appears as a consequence of the $1/\epsilon$
divergence in the $\rho, x_0$ integration. The physical information
contained in the two-point function is in the contribution to the
matrix of anomalous dimensions which is read from the coefficient in
(\ref{4imp-2pt-as}) and does not depend on $\Lambda$.
Unlike the two-point functions
of two impurity operators (\ref{4imp-2pt-as}) is independent of
$\lambda^\pp$, apart from the dependence in the exponential instanton
weight.   The mode-number dependence in (\ref{4imp-2pt-as}) is extremely simple, given the
very large number of terms that had to be summed.

The calculation presented here is not sufficient to determine the
actual instanton induced anomalous dimension of the operator
$\scrO_\mb1$. This requires the diagonalisation of the matrix of
anomalous dimensions of which we have not computed all the entries.
Other entries are determined by the corresponding two-point functions
whose calculation follows the same steps described here and results in
expressions similar to (\ref{4imp-2pt-as}). From this we can conclude
that the behaviour of the leading instanton contribution to the
anomalous dimensions of singlet operators is
\be
\g_\mb1^{\rm inst} \sim  (g_2)^{7/2}
\er^{-\frac{8\pi^2}{g_2\lambda^\pp}+i\theta}
\fr{m_1m_2n_1n_2 } \,
\label{singletadim}
\ee
which is in agreement with the string result
(\ref{finfour}). It is worth stressing that the
the condition of pairwise equality of mode numbers appears in a highly
non-obvious manner in the gauge theory calculation, while it followed
rather trivially from the form of the boundary state in the plane-wave string theory.

The Yang--Mills instanton contributions to other (non-singlet)
four-impurity operators are suppressed by powers of $\lambda'$, as
in the two-impurity case.  This is also in qualitative agreement with the string side of the
correspondence.  However,  in order to evaluate the semi-classical profiles
of the BMN operators we would again have to use the contribution to some of the scalar fields
that contains a product of six fermionic moduli, which presents the same technical obstacle as
in the two impurity case.

\section{Other issues}
\label{concl}

The basic
message is that we find striking agreement between instanton
effects in the gauge theory and those calculated in the plane-wave
string theory.
We focused on operators with two and four scalar impurities since these are the
easiest to calculate on the gauge side. The four
impurity case, although more involved, is fully under control, whereas
the two impurity case presents subtleties due to the fact the leading
semi-classical approximation vanishes and the first non-zero
contribution arises at higher order.
Clearly it would be interesting but very challenging  to generalize
the present work from the one-instanton sector to multi-instanton
sectors.

The structure of the string theory side of the calculation was much simpler than
the gauge side.  In fact, many properties of the Yang--Mills side
would be very difficult to calculate without the insights provided by the string
calculation.  For example, one generic feature of the string
calculation is that only states with an even number of non-zero mode
insertions receive $D$-instanton corrections. Zero mode oscillators can
appear in odd numbers with the condition that they be contracted into
a SO(4)$\times$SO(4) scalar between the incoming and outgoing
states.
Another peculiarity observed in the string theory calculation
is that the $D$-instanton contribution to the masses of
certain states with a large number of fermionic non-zero mode
excitations involves large powers of the mass parameter $m$. These mass-matrix elements
are ones that receive no perturbative contributions.
When expressed in terms of gauge theory parameters this behaviour corresponds to
large {\it inverse} powers of $\lambda^\pp$. This is not pathological in the
$\lambda^\pp\to 0$ limit, because the inverse powers of $\lambda^\pp$
are accompanied by the instanton factor,
$\exp\left(-8\pi^2/g_2\lambda^\pp\right)$. From the point of view of
the gauge theory this result is  intriguing, not only
because of the unusual coupling constant dependence that the anomalous
dimensions of the dual operators display, but also because
there are no other known examples of operators in $\calN$=4 SYM whose
anomalous dimension receives instanton but not perturbative
corrections. This particular class of BMN operators will be discussed in \cite{gks3}.

Finally, we should note that the issue of non-perturbative corrections
to anomalous dimensions is very far removed from the interesting issues surrounding
the integrability of string theory in $AdS_5\times S^5$.  Integrability is
expected to be a property of tree-level string theory and the corresponding planar
approximation to $\calN=4$ Yang--Mills, which can be successfully modelled by
local spin chains.  In contrast, an instanton affects all the fields in the BMN operator
equally, and is therefore highly non-local along the chain.  However, instantons are crucial
in describing the SL(2,$Z$) $S$-duality transformations of the theory and, in
particular, for understanding how SL(2,$Z$) acts on the anomalous dimensions.
In general SL(2,$Z$) transformations relate operators of
small and large dimension, just as in string theory they relate
fundamental strings to $D$-strings, which have large masses of order
$1/g_s$, in the limit of weak string coupling, $g_s\ll 1$. It would be
interesting to understand how $S$-duality is realised in type IIB
string theory in the plane-wave background. A corresponding symmetry
should exist in the BMN sector of $\calN$=4 SYM and the instanton
effects which we have described  should be relevant to its
implementation.

\acknowledgments{AS acknowledges financial support from PPARC and
Gonville and Caius college, Cambridge. We also wish to acknowledge
support from the European Union Marie Curie Superstrings Network
MRTN-CT-2004-512194.}

\end{document}